\documentclass[letterpaper]{article} 
\usepackage[submission]{aaai2026}  
\usepackage{times}  
\usepackage{helvet}  
\usepackage{courier}  
\usepackage[hyphens]{url}  
\usepackage{graphicx} 
\usepackage{natbib}  
\usepackage{caption} 
\usepackage{booktabs}
\usepackage{algorithm}
\usepackage{algorithmic}
\usepackage{amsmath}
\usepackage[utf8]{inputenc}

\usepackage{newfloat}
\usepackage{listings}
\DeclareCaptionStyle{ruled}{labelfont=normalfont,labelsep=colon,strut=off} 
\floatstyle{ruled}
\newfloat{listing}{tb}{lst}{}
\floatname{listing}{Listing}
\title{Bridging Foundation Models and Efficient Architectures: A Modular Brain Imaging Framework with Local Masking and Pretrained Representation Learning}
\author{
    Written by AAAI Press Staff\textsuperscript{\rm 1}\thanks{With help from the AAAI Publications Committee.}\\
    AAAI Style Contributions by Pater Patel Schneider,
    Sunil Issar,\\
    J. Scott Penberthy,
    George Ferguson,
    Hans Guesgen,
    Francisco Cruz\equalcontrib,
    Marc Pujol-Gonzalez\equalcontrib
}
\affiliations{
    \textsuperscript{\rm 1}Association for the Advancement of Artificial Intelligence\\

    1101 Pennsylvania Ave, NW Suite 300\\
    Washington, DC 20004 USA\\
    proceedings-questions@aaai.org
}

\usepackage{graphicx}
\graphicspath{{images/}}
\usepackage{bibentry}
\usepackage{amssymb}
\makeatletter

\def\@maketitle{%
  \newpage \null \vskip 2em%
  \begin{center}%
  {\LARGE\bf \@title \par}%
  \vskip 1.5em%
  {\large                
    \begin{tabular}[t]{c}%
     Yanwen Wang\textsuperscript{1},%
      \;Xinglin Zhao\textsuperscript{1},%
      \;Yijin Song\textsuperscript{1},%
      \;Xiaobo Liu\textsuperscript{2},%
      \;Yanrong Hao\textsuperscript{1},
      \;Rui Cao\textsuperscript{1},%
      \;Xin Wen\textsuperscript{1}\thanks{Corresponding author: Xin Wen, email:xwen@tyut.edu.cn}
    \end{tabular}\par}%
  \vskip 1em%
  {\small
    \textsuperscript{1}Taiyuan University of Technology \\
    \textsuperscript{2}Montreal Neurological Institute Mcgill University \\
    \texttt{\{Yanwen Wang\}2024511304@link.tyut.edu.cn} \\
    \texttt{\{Xin Wen\}xwen@tyut.edu.cn} 
  }
  \end{center}%
  \vskip 1.5em}
\makeatother

\begin{document}
\maketitle

\begin{abstract}
Functional connectivity (FC) derived from resting-state fMRI plays a critical role in personalized predictions such as age and cognitive performance. However, applying foundation models(FM) to fMRI data remains challenging due to its high dimensionality, computational complexity, and the difficulty in capturing complex spatiotemporal dynamics and indirect region-of-interest (ROI) interactions. To address these limitations, we propose a modular neuroimaging framework that integrates principles from FM with efficient, domain-specific architectures. Our approach begins with a Local Masked Autoencoder (LMAE) for pretraining, which reduces the influence of hemodynamic response function (HRF) dynamics and suppresses noise. This is followed by a Random Walk Mixture of Experts (RWMOE) module that clusters features across spatial and temporal dimensions, effectively capturing intricate brain interactions. Finally, a state-space model (SSM)-based predictor performs downstream task inference. Evaluated on the Cambridge Centre for Ageing and Neuroscience (Cam-CAN) dataset, our framework achieved mean absolute errors (MAEs) of 5.343 for age prediction and 2.940 for fluid intelligence, with Pearson correlation coefficients (PCCs) of 0.928 and 0.887, respectively—outperforming existing state-of-the-art methods. Visualization of expert distribution weights further enhances interpretability by identifying key brain regions. This work provides a robust, interpretable alternative to LLM-based approaches for fMRI analysis, offering novel insights into brain aging and cognitive function.
\end{abstract}
\section{Introduction}
Neuroimaging-based prediction of cognitive abilities and brain age holds significant scientific and clinical value, offering insights into cognition development, aging, and clinical outcomes\cite{preimp,metamatching,predis}. A central challenge lies in modeling the complex spatiotemporal patterns inherent in functional brain activity. Functional magnetic resonance imaging (fMRI), provides a non-invasive investigation of brain activity through various derived metrics including amplitude of low-frequency fluctuations (ALFF), regional homogeneity (ReHo), functional connectivity (FC), and dynamic functional connectivity (dFC) which possesses the potential to reveal the mechanisms underlying brain functional development and cognition. 


By employing sliding time windows to model temporal dynamics and region-of-interest (ROI)-based parcellations to reduce spatial complexity, dFC effectively compresses data dimensionality while preserving critical spatiotemporal dependencies. Beyond dimensionality reduction, dFC computation expands the effective sample size by up to 100× through sliding-window segmentation. This is critical for public fMRI datasets, which often have fewer than 1,000 subjects, enabling more effective pretraining of self-supervised models like Masked Autoencoders (MAEs) and improving model generalizability.


Recent advances have seen deep learning models dominate this domain, broadly categorized by input modality: (1) static FC matrices processed by CNNs or GNNs, and (2) raw BOLD time series or segmented dFC sequences modeled using architectures such as Transformers, LSTMs, RNNs, and state-space models (SSMs). More recently, foundation models—particularly large language models (LLMs)—have inspired new downstream frameworks for neuroimaging prediction. However, direct application of LLMs to fMRI data faces key challenges: scanner-specific acquisition protocols introduce temporal misalignment requiring correction, and the hemodynamic response function (HRF) induces inherent delays in BOLD signals, both of which add noise and distort temporal dynamics.

To address these limitations, we propose a modular brain imaging framework that integrates principles from foundation models with efficient, fMRI-specific architectural designs. Our goal is to effectively capture complex spatiotemporal dependencies and indirect ROI interactions while handling high-dimensional data efficiently. The key contributions are threefold: 
\begin{itemize}
    \item We propose a novel modular framework incorporating Local Masked Autoencoding (LMAE) and pretrained representation learning, effectively adapting large language model (LLM)-inspired architectures to fMRI data for brain age and cognitive ability prediction.
    
    \item We improve model interpretability by analyzing the expert distribution within the Random Walk Mixture of Experts (RWMOE) module from the perspective of functional brain networks, revealing neuroscientifically meaningful patterns.
    
    \item We evaluate our method on the publicly available Cam-CAN dataset, demonstrating significant performance improvements over state-of-the-art approaches in predicting both age and fluid intelligence.
\end{itemize}
\section{Related Work}
\subsection{fMRI-based Predictive models}

Over the past decade, a growing body of research has leveraged deep learning to model FC patterns\cite{Alterations}, enabling more accurate and interpretable brain-behavior predictions\cite{PENG}.

GNNs have emerged as a dominant paradigm in brain network analysis, owing to their ability to model irregular topological structures. BrainGNN~\cite{braingnn} is a pioneering framework that identifies task- or disease-relevant brain regions by integrating graph convolution with ROI selection. Similarly, ST-DAG-Att~\cite{ST-DAG} models the brain as a dynamic graph, with nodes representing ROIs and edges capturing functional connectivity\cite{Plastic}, while incorporating structural priors into the graph architecture. To better exploit dense connectivity patterns, Graph Path Convolution~\cite{GPC} introduces an edge-based message-passing mechanism that effectively propagates information across richly connected brain networks.

In parallel, attention mechanisms have been widely adopted to enhance model expressiveness and interpretability. The Multi-Level Joint Attention Network~\cite{ML-Joint-Att} learns hierarchical FC representations by combining node and edge level convolutions with adaptive Inception modules and multi-scale attention, enabling joint modeling across cognitive tasks. The Global-Local Brain Network~\cite{glbn} improves prediction of age and fluid intelligence by fusing whole-brain context with salient local connectivity patterns through a dual-branch attention architecture.

More recently, Transformer-based models have been introduced to capture long-range spatiotemporal dependencies in fMRI data. SwiFT (Swin 4D fMRI Transformer)~\cite{swin} adopts a 4D patch-based Transformer to learn dynamic brain features directly from fMRI volumes in an end-to-end manner, bypassing the need for pre-defined parcellations. Despite these advances, most existing methods rely on hand-crafted architectures or static representations, and none have effectively integrated principles from foundation models into fMRI representation learning\cite{resnet,systematic}, leaving significant room for improvement in both performance and generalization.

\subsection{FM in Neuroimaging}
Large-scale models in neuroimaging have predominantly focused on decoding tasks using electroencephalography (EEG) data, where the temporal resolution is high and the signals are more directly related to neural activity~\cite{eegdecoding,eegllm,nipseeg}. These models leverage the rich temporal dynamics of EEG for various applications such as emotion recognition, cognitive state monitoring, and motor imagery classification. However, their application to fMRI-based prediction tasks remains largely unexplored.

Despite significant advancements in deep learning techniques for fMRI analysis, including CNNs, GNNs, transformers, and RNNs~\cite{medicalimage,braingnn,brainaging}, large language models (LLMs) have yet to be effectively applied to fMRI data. This gap can be attributed to a fundamental challenge: extracting meaningful semantic information from fMRI data remains difficult due to its complex spatiotemporal structure and the inherent noise~\cite{taskrs}. Although there have been some preliminary attempts to integrate LLMs with fMRI data\cite{llmfnc}, these efforts have not achieved predictive performance that surpasses recent conventional deep learning models.

Therefore, exploring how to effectively adapt LLMs for fMRI semantic representation and apply them to downstream predictive tasks represents a critical research direction. Addressing this challenge could lead to breakthroughs in understanding brain aging and cognitive functions, offering new insights and methodologies for neuroimaging analysis.
\section{Methodology}
\subsection{Overall Architecture}
Figure 1 presents the overall architecture of the proposed model. The framework begins with a Local Mask Autoencoder (LMAE) for hierarchical feature extraction. The LMAE encoder processes only partially masked input data—strategically omitting segments to simulate neural response delays—thereby mapping the observed fMRI time series into a compact latent representation. This design explicitly accounts for the maximum delay in brain responses, reducing sensitivity to HRF variability. Concurrently, it enables dimensionality reduction of FC matrices computed via Pearson correlation across ROI time series. A lightweight decoder reconstructs the full input during pretraining; however, it is discarded in downstream tasks, with only the encoder retained for feature extraction.

The extracted latent features are then processed by a Random Walk Mixture of Experts (RW-MoE) module, which performs structured clustering along both spatial and temporal dimensions. By leveraging a random walk mechanism, RW-MoE adaptively groups functionally coherent brain regions and temporal patterns, enhancing representational expressiveness.

Finally, a predictor head based on the Mamba architecture generates the output. Built upon a Selective State Space Model and augmented with an MLP subnetwork, this component efficiently captures long-range temporal dependencies and instantaneous dynamics in fMRI signals, while exhibiting strong robustness to noise—a critical advantage in neuroimaging applications.
\begin{figure*}[htb]
    \centering
    \includegraphics[width=0.9\textwidth, height=0.4\textheight]{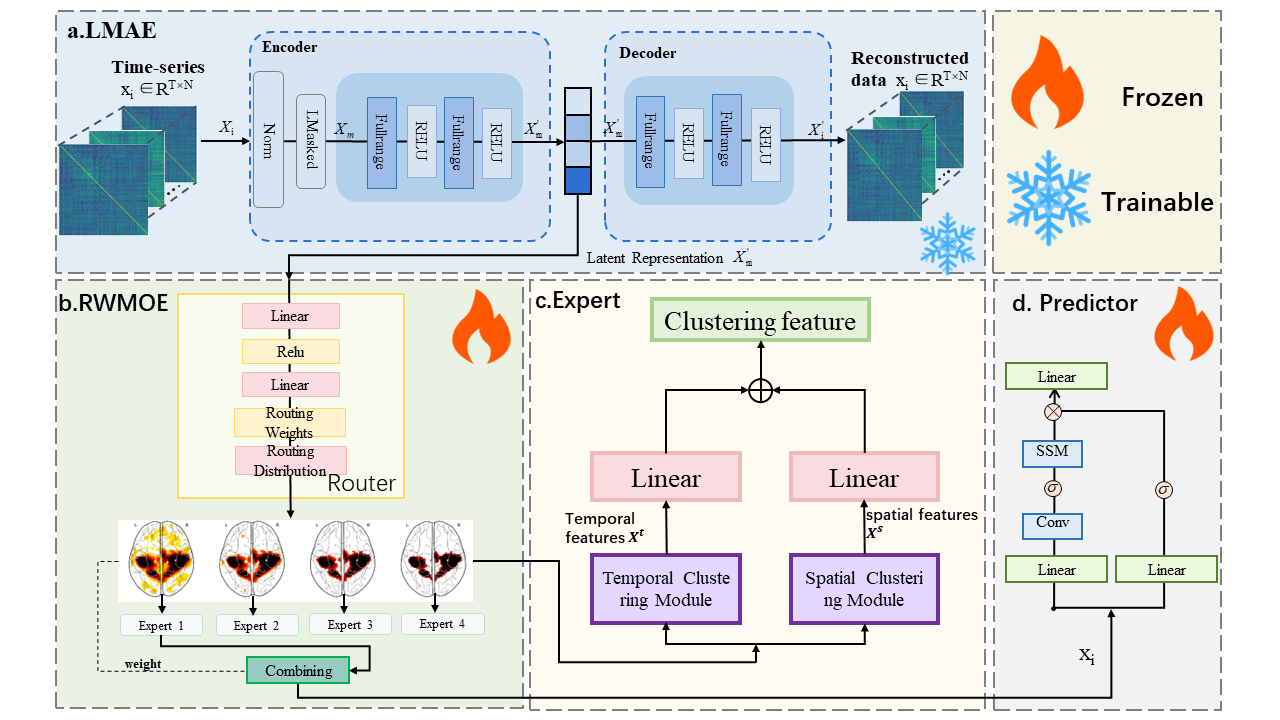}
    \caption{Overall Architecture:Firstly, (a)LMAE is used for feature extraction;(b) RWMOE is used for feature clustering from time dimension and space dimension respectively;(c)Expert Model:Double branch structure is used to cluster features from time dimension and space dimension;(d) a predictor is used to predict the results.}
    \label{fig:model_architecture}
\end{figure*}
\subsection{Local Masked Autoencoders}
In current research, LMAE is proposed as a refined self-coding method. Its core goal is to reconstruct the complete output data through partial original input data, and obtain the latent representation of the input data mapping as the extracted feature to solve the problem that the FC data obtained by calculating the Pearson correlation coefficient of the original data are difficult to process due to the large dimension. Compared with the classical autoencoder, LMAE introduces a unique mask mechanism, which significantly improves the performance and efficiency of the model. 
\begin{figure}[htb]
    \centering
    \includegraphics[width=0.5\textwidth, height=0.3\textheight]{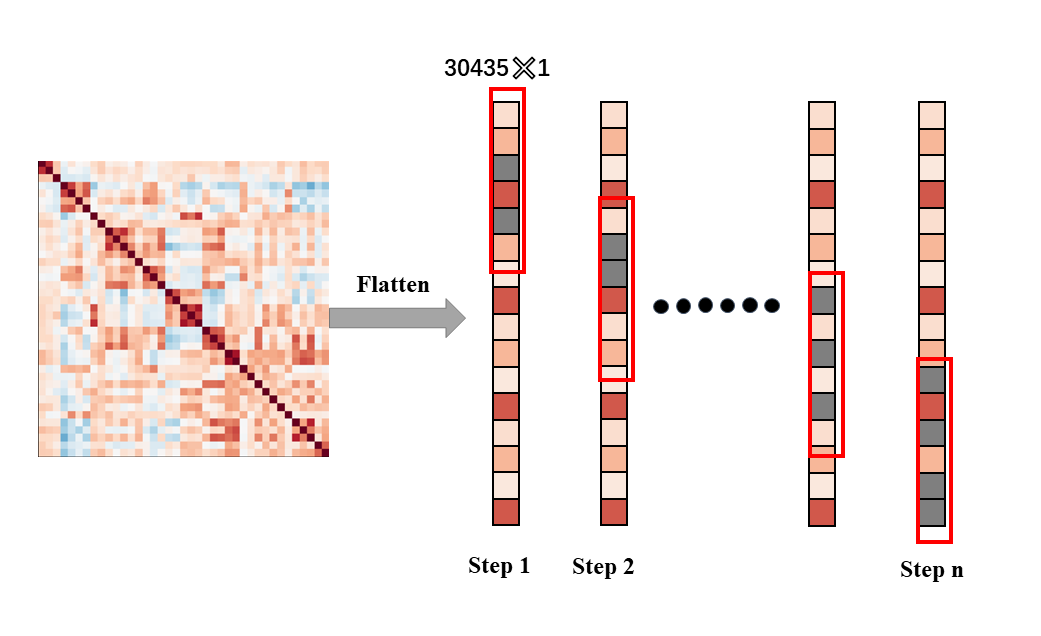}
    \caption{Mask Strategy: The local region random mask mechanism.}
    \label{fig:LMAE}
\end{figure}
\subsubsection{Mask Strategy.} 
In classical autoencoders, the entire input is fed to the encoder. While this design is simple, it may suffer from high computational cost and low reconstruction efficiency when processing large-scale data. In contrast, LMAE introduces a local region masking mechanism. Unlike traditional discrete patch masking, LMAE divides the input into contiguous windows, where masking is applied to continuous local segments. This windowing strategy accounts for the maximum delay in brain response, as illustrated in Figure 2, enabling more accurate capture of the temporal characteristics of brain activity and HRF dynamics. By setting an appropriate mask ratio, data redundancy can be significantly reduced. The specific formula is as follows:

\begin{equation}
Z_{i,j} = \frac{X_{i,j} - \mu}{\sigma}
\end{equation}
\begin{equation}
X_{i,j}^{\text{norm}} = 2 \times \frac{Z_{i,j} - \min(Z_{i,j})}{\max(Z_{i,j}) - \min(Z_{i,j})} - 1
\end{equation}
\begin{equation}
i \sim \text{Uniform}(0, N-b)
\end{equation}
\begin{equation}
M[:, i:i+b] = 0_{T \times b}
\end{equation}
\begin{equation}
X_{i,j}^{m} = X_{i,j}^{\text{norm}} \odot M
\end{equation}
Let \( X_{i,j} \in \mathbb{R}^{T \times N} \) be the input data, where \( T \) represents the number of time points and \( N \) represents the number of Regions of Interest (ROIs). \( X_{i,j} \) denotes the value at the \( i \)-th time point and the \( j \)-th ROI. Before applying the mask, the data is first standardized, where \( \mu \) is the global mean and \( \sigma \) is the global standard deviation. The standardized data is denoted as \( Z_{i,j} \), and the normalized data is denoted as \( X_{i,j}^{\text{norm}} \). The mask matrix is denoted as \( M \in \mathbb{R}^{T \times N} \), and the block size parameter of the mask is \( b \), defined as \( b = (\alpha \times D) \), where \( \alpha \) is the mask ratio and \( D \) is the size of the input data. The randomly generated index matrix is denoted as \( i \), and the masked data is denoted as \( X_{i,j}^{m} \).
\subsubsection{Encoder.} During input processing, the encoder focuses exclusively on the unmasked regions and completely disregards the masked portions. This design enables the encoder to concentrate on available information, avoiding redundant computations on masked data and thereby reducing computational complexity. The encoder, composed of two one-dimensional convolutional layers followed by a fully connected layer, performs feature extraction through stepwise downsampling. It maps the input sequence to a low-dimensional latent space, producing compact representations that the subsequent decoder utilizes to reconstruct the original input.
\subsubsection{Decoder.} 
During reconstruction of the original input, the decoder utilizes the latent representation generated by the encoder to efficiently recover the complete data. The decoder employs a lightweight architecture for fast processing, consisting of two deconvolutional layers that progressively upsample the latent features to match the original input dimension. Finally, a $\tanh$ activation function is applied to constrain the output values within the interval  $(-1, 1)$, ensuring stable and high-quality reconstruction.

The decoder is only used to perform the reconstruction task of the input data during pre-training. The loss function calculates the mean square error (MSE) of the reconstructed data and the original input data.
\begin{equation}
L_R = (X^{\text{dec}}, \alpha_i, L_i) = \frac{1}{n} \sum_{i=1}^{n} \left\| X^{\text{dec}} - X_{i,j}^{\text{norm}} \right\|^2
\end{equation}
\( n \) is the number of samples, \( \alpha_i \) is the parameter of the masking ratio, and \( L_i \) is the parameter of the dimension of encoded features.
\subsection{Random Walk Mixture of Experts}

After obtaining the features extracted by the encoder, the Random Walk Mixture of Experts dynamically routes the input features to the most suitable expert via a distributed router. Each expert specializes in processing specific patterns in the data and performs feature clustering along both the temporal and spatial dimensions, addressing the limitation of standard feedforward networks in modeling complex cross-temporal variable interactions. The outputs of all activated experts are then aggregated to produce the final representation. This mechanism promotes balanced expert utilization, enhances modeling efficiency, and improves the model’s generalization capability. As illustrated in Figure 1.
\subsubsection{Distributed Router.}The router is a simplified neural network model, and the input is the feature vector extracted by the encoder. The specific formula is as follows :
\begin{equation}
\text{Router}(X^{\text{enc}}) = \text{ReLU}(X^{\text{enc}} \times W_0) \times W_1
\end{equation}
\begin{equation}
H(X^{\text{enc}}) = W^H \times \text{Router}(X^{\text{enc}})
\end{equation}
$X^{\text{enc}}$ is the encoded data, $W_0$ and $W_1$ are the weights of the router, and $W^H$ is the weight of the distribution projection. $H(X^{\text{enc}})$ represents the distribution projection. During training, the network learns the correlation between the input data and each expert, calculating a weight for each expert. The router uses these calculated weights to assign the input data to the corresponding expert node for processing, thus completing the routing task.

In the design of routers, in order to prevent some experts from being selected too frequently and other experts from not being trained enough, the noise mechanism is introduced. The specific method is to add a certain amount of noise to the output logits of the router, and then apply the softmax function.
\subsubsection{Top-K Selection Strategy.} In addition to the weighted sum method based on softmax weights, routers can also use Top-K selection strategy to further simplify calculations and improve efficiency. The strategy selects the experts with the top K weights by parameterization for subsequent calculations, while ignoring the experts with lower weights. The specific formula is as follows :  
\begin{equation}
G(X^{\text{enc}}) = \text{Softmax}(\text{TopK}(H(X^{\text{enc}}), k))
\end{equation}
\begin{equation}
\text{TopK}(H(X^{\text{enc}}), k)_i = 
\begin{cases} 
H(X^{\text{enc}})_i &\text{if }i\in\text{ArgTopK}(H) \\
-\infty &\text{otherwise}
\end{cases}
\end{equation}
$G(X^{\text{enc}}) \in \mathbb{R}^k$ represents the probability of each candidate distribution, H is $(H(X^{\text{enc}}))$ represents the distribution projection and $\text{TopK}(H(X^{\text{enc}}), k)$ represents the top $k$ most likely potential distributions.
\subsubsection{Expert Model.}  Double branch structure is used to cluster features from time dimension and space dimension. As shown in Figure 1.
\subsubsection{Time Clustering Module.} The structure of Feed-Forward Neural Network ( FFN ) is used to enable the model to capture patterns of complex time features. The specific formula is as follows :
\begin{equation}
X^{\text{temp}}(X^{\text{enc}}) = \text{Linear}(\text{RELU}(X^{\text{enc}}))
\end{equation}
$X^{\text{enc}}$ is the encoded data, and $X^{\text{temp}}$ is the temporal feature generated by the time clustering module.
\subsubsection{Spatial Clustering Module.} The limitation of using the original graph neural network for brain network analysis is that it cannot model the interaction between indirectly connected neighbors, because it can only infer direct pairwise connections between brain ROIs. In this case, the interaction between multiple brain regions or brain regions indirectly connected through the intermediate region may be ignored. However, broader node interactions between these indirectly connected brain regions are also crucial. In order to incorporate this broader interaction into the proposed method, the random walk process is further applied to the spatial clustering module.  

The structure of random walk graph neural network is adopted, and the core of this method is random walk operation. Specifically : graph propagation through the adjacency matrix A. At each step of propagation, the feature vector of the node is multiplied by the adjacency matrix to update the node features according to the structure of the graph. In the loop, \text{walk\_depth} subgraph propagation is performed to update the feature vector. The specific formula is as follows : 
\begin{equation}
X^{(\text{walk\_depth})} = A^{(\text{walk\_depth})} \times X^{\text{enc}}
\end{equation}
\begin{equation}
X^{\text{spat}}(X^{\text{enc}}) = \text{Linear}(\text{ReLU}(\text{Linear}(X^{(\text{walk\_depth})})))
\end{equation}
Where \( A \in \mathbb{R}^{N \times N} \) is the adjacency matrix of the graph, representing the connection relationships between nodes in the graph. \( X^{\text{enc}} \in \mathbb{R}^{T \times N} \) are the features extracted by the LMAE. \( \text{walk\_depth} \) is the parameter for the number of steps in the random walk, and \( X^{(\text{walk\_depth})} \) represents feature propagation on the graph structure, as shown in Algorithm 1. \( X^{\text{spat}} \) is the spatial feature generated by the spatial clustering module. By combining the random walk strategy with graph convolution, complex spatial feature patterns can be fully captured, allowing for the adaptive extraction of more extensive dependencies between brain ROIs. These features can better reveal the complexity and scope of interactions in brain networks.
\begin{algorithm}
\caption{Graph propagation using random walk mechanism}
\label{alg:simplified_gcn}
\begin{algorithmic}[1]

\STATE \textbf{Input:}
\STATE \quad $A \in \mathbb{R}^{D \times D}$: Adjacency matrix 
\STATE \quad $X \in \mathbb{R}^{B \times T \times D}$: Input features
\STATE \quad $\text{walk\_depth} \in \mathbb{N}$: Number of propagation steps
\STATE \textbf{Output:}
\STATE \quad $X^{(\text{walk\_depth})} \in \mathbb{R}^{B \times T \times D}$: Walk-processed features

\STATE \textbf{Step 1: Initialize Propagation}
\STATE $X^{(0)} \leftarrow X$ \hfill 
\STATE $X^{(\text{current})} \leftarrow X$ \hfill 
\STATE $H \leftarrow \text{zeros}(B \times T \times D)$ \hfill 

\STATE \textbf{Step 2: Multi-step Feature Propagation}
\FOR{$\text{step} = 1$ \TO $\text{walk\_depth}$}
    \FOR{each batch $b$ and time step $t$}
        \STATE $X^{(\text{current})}_{b,t,:} \leftarrow X^{(\text{current})}_{b,t,:} \cdot A^T$ \hfill 
    \ENDFOR 
    \STATE $H \leftarrow H + X^{(\text{current})}$
    
    \IF{$\text{step} < \text{walk\_depth}$}
        \STATE $X^{(\text{current})} \leftarrow \text{ReLU}(X^{(\text{current})})$
    \ENDIF
\ENDFOR

\STATE \textbf{Step 3: Final Output}
\STATE $X^{(\text{walk\_depth})} \leftarrow H + X^{(0)}$ \hfill 
\STATE \textbf{return} $X^{(\text{walk\_depth})}$

\end{algorithmic}
\end{algorithm}

Finally, the time feature and the spatial feature are connected in series to obtain the fusion feature, which takes into account the complex changes of the time dimension and the spatial dimension. The specific formula is as follows :
\begin{equation}
X^{\text{mix}} = \text{Linear}(\text{concat}(X^{\text{temp}}, X^{\text{spat}}))
\end{equation}
$X^{\text{mix}}$ is a fusion feature of temporal features and spatial features.
\subsubsection{Combination Function.} It is responsible for synthesizing the output of multiple expert models. Each expert model focuses on specific input features and obtains the final result by weighted average to ensure that the output of each expert is reasonably valued. 
\subsection{Predictor}
The predictor is a Mamba framework with SSM architecture design and Transformers ' MLP module, which can effectively capture the instantaneous dynamics in time series data and improve the robustness of the model to data noise problems.

The core innovation of the Mamba model is the introduction of a selectivity mechanism, which enables the model to dynamically adjust its behavior according to different inputs by making the SSM parameters dependent on the input. In order to optimize the computational efficiency, Mamba adopts a hardware-aware algorithm, especially using the GPU 's memory hierarchy to improve the calculation speed of the scanning operation and reduce the memory requirements. This method combines the recursive computing efficiency of RNN and the parallel processing advantages of CNN, making Mamba more efficient in processing long sequence data.The specific formula is as follows :
\begin{align}
M_l &= \text{Mamba}(X_{l-1}^{\text{feat}}; \theta_l)
\end{align}
$M_l$ represents the output of the $l$-th mamba block, $\theta_l$ represents the parameters of the $l$-th mamba block, and $X_{l-1}^{\text{feat}}$ represents the features processed by the $l-1$-th layer.
\begin{equation}
X_{\text{avg}} = \frac{1}{L} \sum_{t=1}^{L} X_l^{\text{feat}}[:, t, :]
\end{equation}
$X_{\text{avg}}$ represents the sequence aggregation feature, that is, the global feature of each sample.
\begin{equation}
Y = W_{\text{reg}} \times \text{LayerNorm}(X_{\text{avg}}) + b_{\text{reg}}
\end{equation}
where $Y$ is the output result, $W_{\text{reg}}$ and $b_{\text{reg}}$ are the weight and bias of the regression head, respectively.
\section{Experiments}
\begin{table*}[t]
\centering
\begin{tabular}{ccccc}
\toprule
\textbf{Models} & \textbf{FI MAE} & \textbf{FI PCC} & \textbf{Age MAE} & \textbf{Age PCC} \\ 
\midrule
Vision-mamba (2024) & 4.312 $\pm$ 0.233 & 0.620 $\pm$ 0.039 & 7.291 $\pm$ 0.742 & 0.916 $\pm$ 0.029 \\
S-Mamba (2024) & 4.431 $\pm$ 0.461 & 0.622 $\pm$ 0.038 & 7.312 $\pm$ 0.556 & 0.891 $\pm$ 0.014 \\
iTransformer (2024) & 4.260 $\pm$ 0.153 & 0.765 $\pm$ 0.087 & 7.556 $\pm$ 0.304 & 0.892 $\pm$ 0.023 \\
Swin-Transformer (2021) & 4.562 $\pm$ 0.203 & 0.789 $\pm$ 0.094 & 6.934 $\pm$ 0.355 & 0.897 $\pm$ 0.021 \\
FLDMamba (2025) & 4.125 $\pm$ 0.261 & 0.815 $\pm$ 0.097 & 5.840 $\pm$ 0.803 & 0.919 $\pm$ 0.032 \\
Ours & 2.940 $\pm$ 0.251 & 0.887 $\pm$ 0.067 & 5.343 $\pm$ 0.352 & 0.928 $\pm$ 0.036 \\
\bottomrule
\end{tabular}
\caption{Comparison of MAE and PCC values across different models for Cam-CAN dataset.}
\end{table*}
\subsection{Experimental Setup}
\subsubsection{Datasets.} Cambridge Centre for Ageing and Neuroscience ( Cam-CAN ) is a large-scale cognitive aging research initiative. We use the fMRI data of this dataset.
\subsubsection{Preprocessing.} The pre-processing uses fMRIPrep pre-processing pipeline to preprocess the original image, and the brain network group map is used to partition the brain. The Pearson correlation coefficient between the ROIs time series was calculated to obtain the dFC.
\subsubsection{Implementation Details.} The proposed model is based on the PyTorch 2.1.1, uses a five-fold cross-validation strategy, and uses a five-fold average MAE to evaluate the performance of the model. Configure the batch size to 64 and use the Adam optimizer. The initial learning rate is set to 1e-2, and the loss function is L1Loss.
\subsection{Results} In order to evaluate the performance of the model, our model is compared with other models. The results of the comparison model are shown in Table 1. The results show that our proposed modular neural imagination framework effectively captures complex spatiotemporal dynamics and indirect ROI interactions, and achieves the best results in prediction tasks. Our method shows significant advantages in both FI (fluid intelligence) and Age (age) prediction tasks.

In the FI prediction task, our method achieves the lowest MAE value of 2.940, which is 1.185 lower than the existing best method FLDMamba (4.125). At the same time, the PCC ( Pearson correlation coefficient ) index also reached the highest 0.887, which was significantly better than all other comparison models. In the Age prediction task, our method achieves the best performance again, with MAE of 5.343, which is significantly better than FLDMamba (5.840) and other models. In addition, we also reached the highest 0.928 in PCC.

\subsection{Ablation Study}
In order to determine the effects of different components, ablation studies were performed on the following parts. The results of ablation experiments are shown in Table 2.
\\( 1 ) W / O LMAE : Removing the local mask autoencoder will significantly reduce the model performance. Specifically, MAE values increased significantly and PCC values decreased significantly in age prediction and fluid intelligence prediction tasks. It shows that the module effectively extracts features.
\\( 2 ) W / O RWMOE : Removing the random walk hybrid expert module will significantly reduce the performance of the model. The MAE values increased from 5.343 to 5.710 and 2.940 to 4.077, respectively. The PCC values decreased from 0.928 to 0.902 and 0.887 to 0.772, respectively. It shows that RWMOE can perform feature clustering from the time dimension and the spatial dimension.
\\( 3 ) W / O local patch : Removing the continuous window partition mechanism, although the PCC value changes relatively small, the MAE value increases significantly. It shows that this component can effectively capture complex spatiotemporal dynamics. 
\\( 4 ) W / O Random Walk : The effect of removing the random walk mechanism is similar to that of Local Patch, indicating that this component can effectively capture indirect region of interest ( ROI ) interactions.
\\( 5 ) W / O local patch and Random Walk : The performance degradation after removing the continuous window partition and random walk mechanism is more obvious than removing the Local patch or Random Walk alone, indicating that the two components have a synergistic effect in capturing complex spatiotemporal dynamics and indirect region of interest ( ROI ) interaction.
\begin{table*}[t]
\centering
\begin{tabular}{lcccccc}
\toprule
\textbf{Models} & \textbf{FI MAE} & \textbf{FI PCC} & \textbf{Age MAE} & \textbf{Age PCC} \\
\midrule
W/O LMAE & 4.182 & 0.779 & 6.191 & 0.866 \\
W/O RWMOE & 4.077 & 0.772 & 5.710 & 0.902 \\
W/O local patch and Random Walk & 3.131 & 0.684 & 5.664 & 0.873 \\
W/O local patch & 3.040 & 0.769 & 5.453 & 0.915 \\
W/O Random Walk & 3.022 & 0.762 & 5.407 & 0.917 \\
Ours & 2.940 & 0.887 & 5.343 & 0.928 \\
\bottomrule
\end{tabular}
\caption{Ablation study of each component.}
\end{table*}
\subsection{Parameter Analysis}
\subsubsection{Mask ratio.} In the LMAE Encoder block, the mask ratio $\alpha$ is a hyperparameter that affects the prediction results. To determine the optimal value, experiments were conducted on prediction tasks with different mask ratios, specifically $\alpha = \{ 0.1, 0.2, \dots, 0.9 \}$. Figure 3 presents the results for each prediction task at different $\alpha$ values. The results show that for the Cam-CAN FI metric prediction task, the optimal $\alpha = 0.2$, which corresponds to the lowest MAE and the highest PCC. For the Cam-CAN Age metric prediction task, the optimal $\alpha = 0.3$, which corresponds to the lowest MAE and the highest PCC. Therefore, the proposed model selects mask ratios of $\alpha = 0.2$ and $\alpha = 0.3$.
\begin{figure}[H]
    \centering
    \includegraphics[width=0.45\textwidth, height=0.3\textheight]{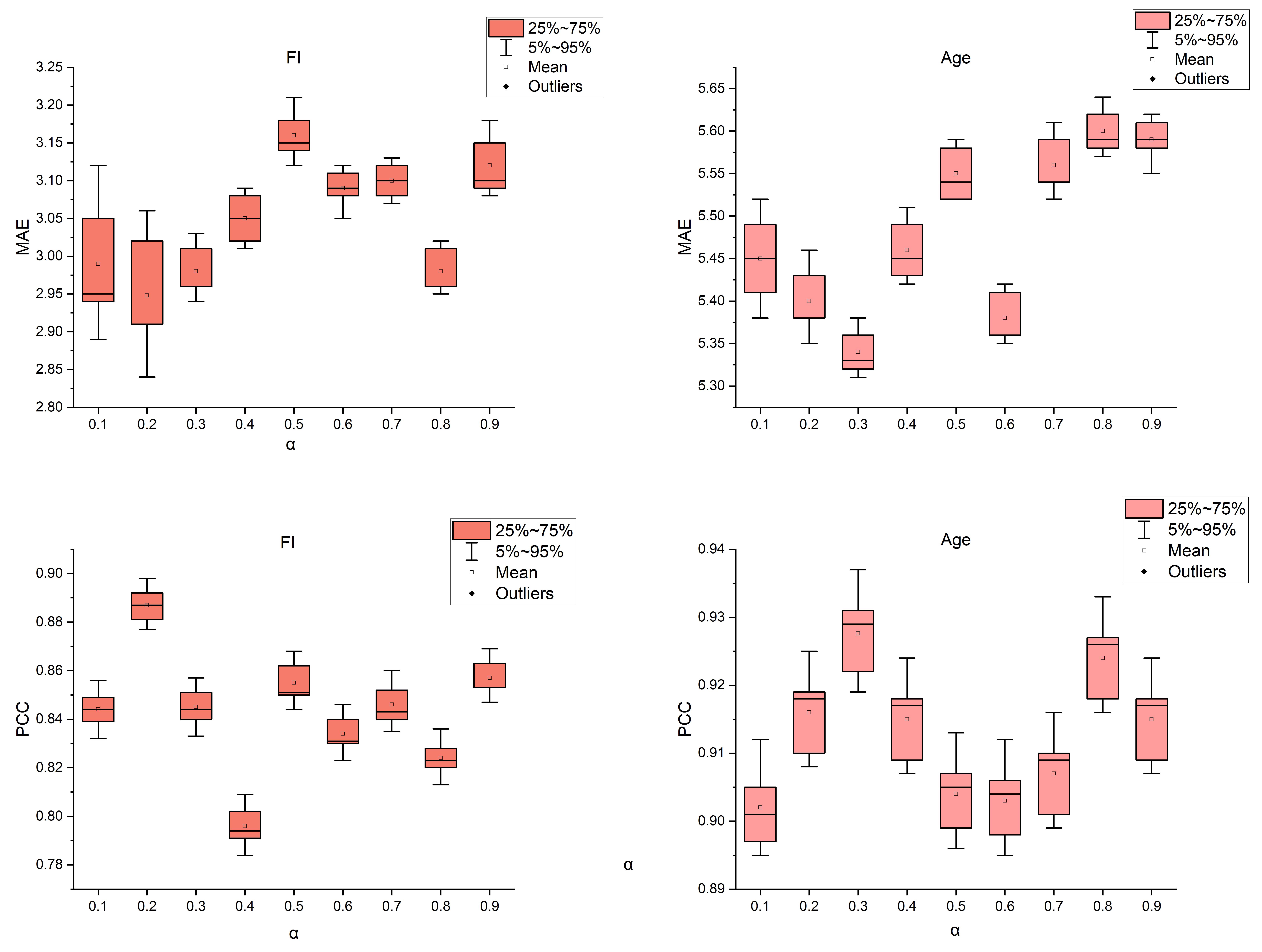}
    \caption{MAEs and PCCs under different mask ratios $\alpha$ are used to predict age and fluid intelligence.}
    \label{fig:Mask ratio}
\end{figure}
\subsubsection{Hiding layer size and the dimension of latent features.}The size of the hidden layer determines the feature dimension on which the routing decision depends. If the dimension is insufficient, it may lead to inaccurate routing allocation and affect the efficiency of expert division of labor. In order to determine the optimal value, experiments are carried out on different hidden layer sizes, specifically \texttt{hidden\_size = \{64,128,256,512\}}. The prediction results show that the optimal hidden layer size is 128, the corresponding MAE is the lowest, and the PCC is the highest.

In LMAE 's Encoder block, the dimension of the latent feature acts as a hyperparameter that affects the prediction results. In order to determine the optimal value, the prediction tasks of different latent feature dimensions are tested, specifically $\{32,62,64,128\}$. The results show that the dimension of the optimal latent feature is 62, and the corresponding MAE is the lowest and the PCC is the highest. As shown in Figure 4.

\begin{figure}[t]
    \centering
    \includegraphics[width=0.3\textwidth, height=0.2\textheight]{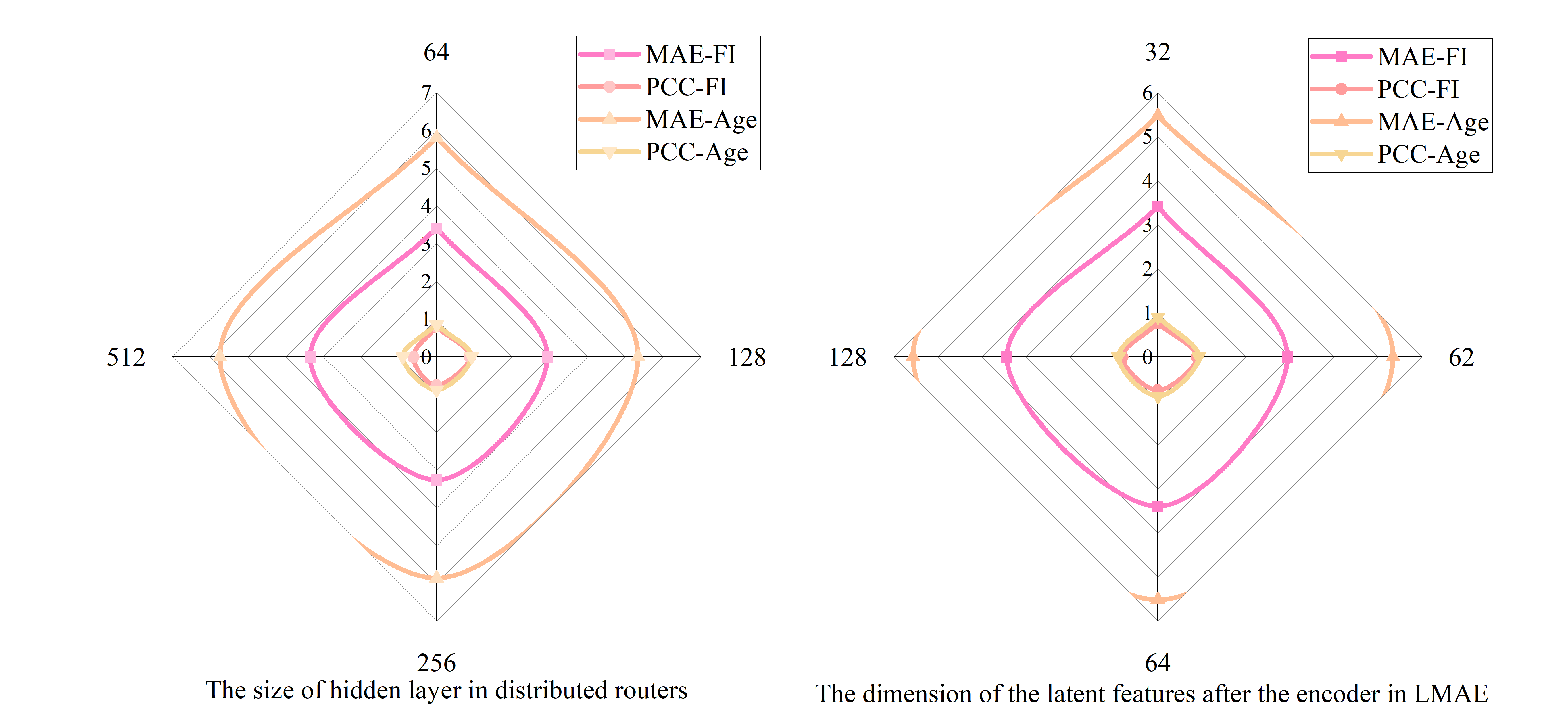}
    \caption{MAEs and PCCs with different hidden layer sizes and the dimension of latent features are used to predict age and fluid intelligence.}
    \label{fig:Mask ratio}
\end{figure}

\subsection{Visual distribution weight}
The number of experts is set to 4, and K=1, meaning only the top-ranked expert is activated for each input. The distributed router in RW-MoE computes a weight for each expert by learning the correlation between the input features and the expert’s specialization. Figure 5 visualizes the expert preference across different ROIs, illustrating the routing behavior of RWMOE. The results show that each expert exhibits distinct regional preferences, with functionally related ROIs clustered under the same expert. This specialization enables the model to capture heterogeneous spatial patterns and enhances prediction performance by promoting expert diversity. The non-uniform assignment of ROIs to experts demonstrates the adaptability of RWMOE in modeling region-specific dynamics, highlighting its capacity for interpretable and structured representation learning.
\begin{figure*}[t]
    \centering
    \includegraphics[width=0.9\textwidth, height=0.6\textheight]{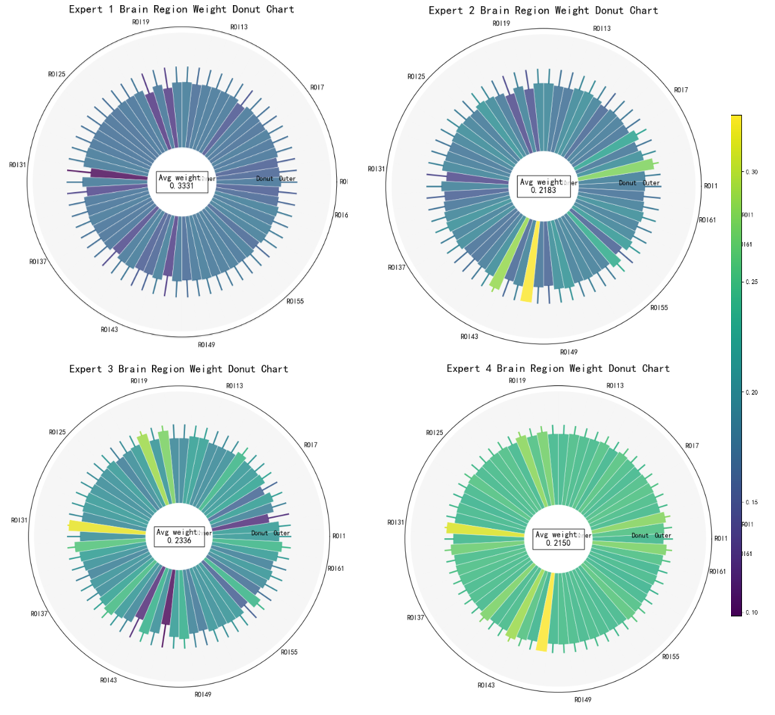}
    \caption{Each expert 's preference for different ROIs.}
    \label{fig:model_architecture}
\end{figure*}
\section{Conclusion}
In this work, we present a novel, modular neuroimaging framework that bridges the gap between foundation model principles and domain-specific efficiency for analyzing high-dimensional fMRI data. By introducing a Local Masked Autoencoder (LMAE) with biologically informed masking, our method effectively reduces sensitivity to hemodynamic response variability and suppresses noise during pretraining. The Random Walk Mixture of Experts (RW-MoE) module further enables adaptive, interpretable clustering of spatiotemporal features, capturing complex interactions across brain regions and time points that are often missed by standard architectures. Finally, the integration of a state-space model predictor ensures efficient and robust inference on downstream tasks. Evaluated on the Cam-CAN dataset, our framework achieves state-of-the-art performance in predicting age and fluid intelligence, with low MAEs and high correlation coefficients, demonstrating its strong representational power and generalization ability. Moreover, the learned expert distributions provide interpretable insights into neurocognitive organization, highlighting functionally relevant brain regions. Unlike generic large language model-inspired approaches, our method offers a lightweight, transparent, and neuroscience-grounded alternative tailored to fMRI analysis. This work not only advances the methodology for brain-behavior modeling but also opens new avenues for interpretable, foundation-model-inspired architectures in medical imaging and cognitive neuroscience.

\bigskip

\bibliography{aaai2026}
\newpage
\clearpage
\section{Reproducibility Checklist}
\paragraph{This paper:}
\begin{itemize}
    \item Includes a conceptual outline and/or pseudocode description of AI methods introduced (yes/partial/no/NA) {\bf yes}
    \item Clearly delineates statements that are opinions, hypothesis, and speculation from objective facts and results (yes/no) {\bf yes}
    \item Provides well marked pedagogical references for less-familiare readers to gain background necessary to replicate the paper (yes/no) {\bf yes}
\end{itemize}

\paragraph{Does this paper make theoretical contributions? (yes/no)} {\bf yes}

If yes, please complete the list below.
\begin{itemize}
    \item All assumptions and restrictions are stated clearly and formally. (yes/partial/no) {\bf yes}
    \item All novel claims are stated formally (e.g., in theorem statements). (yes/partial/no) {\bf yes}
    \item Proofs of all novel claims are included. (yes/partial/no) {\bf yes}
    \item Proof sketches or intuitions are given for complex and/or novel results. (yes/partial/no) {\bf yes}
    \item Appropriate citations to theoretical tools used are given. (yes/partial/no) {\bf yes}
    \item All theoretical claims are demonstrated empirically to hold. (yes/partial/no/NA) {\bf yes}
    \item All experimental code used to eliminate or disprove claims is included. (yes/no/NA) {\bf NA}
\end{itemize}

\paragraph{Does this paper rely on one or more datasets? (yes/no)} {\bf yes}

If yes, please complete the list below.
\begin{itemize}
    \item A motivation is given for why the experiments are conducted on the selected datasets (yes/partial/no/NA) {\bf yes}
    \item All novel datasets introduced in this paper are included in a data appendix. (yes/partial/no/NA) {\bf yes}
    \item All novel datasets introduced in this paper will be made publicly available upon publication of the paper with a license that allows free usage for research purposes. (yes/partial/no/NA) {\bf yes}
    \item All datasets drawn from the existing literature (potentially including authors’ own previously published work) are accompanied by appropriate citations. (yes/no/NA) {\bf yes}
    \item All datasets drawn from the existing literature (potentially including authors’ own previously published work) are publicly available. (yes/partial/no/NA) {\bf yes}
    \item All datasets that are not publicly available are described in detail, with explanation why publicly available alternatives are not scientifically satisficing. (yes/partial/no/NA) {\bf NA}
\end{itemize}

\paragraph{Does this paper include computational experiments? (yes/no)} {\bf yes}

If yes, please complete the list below.
\begin{itemize}
    \item Any code required for pre-processing data is included in the appendix. (yes/partial/no) {\bf no}
    \item All source code required for conducting and analyzing the experiments is included in a code appendix. (yes/partial/no) {\bf no}
    \item All source code required for conducting and analyzing the experiments will be made publicly available upon publication of the paper with a license that allows free usage for research purposes. (yes/partial/no) {\bf yes}
    \item All source code implementing new methods have comments detailing the implementation, with references to the paper where each step comes from. (yes/partial/no) {\bf yes}
    \item If an algorithm depends on randomness, then the method used for setting seeds is described in a way sufficient to allow replication of results. (yes/partial/no/NA) {\bf yes}
    \item This paper specifies the computing infrastructure used for running experiments (hardware and software), including GPU/CPU models; amount of memory; operating system; names and versions of relevant software libraries and frameworks. (yes/partial/no) {\bf yes}
    \item This paper formally describes evaluation metrics used and explains the motivation for choosing these metrics. (yes/partial/no) {\bf yes}
    \item This paper states the number of algorithm runs used to compute each reported result. (yes/no) {\bf yes}
    \item Analysis of experiments goes beyond single-dimensional summaries of performance (e.g., average; median) to include measures of variation, confidence, or other distributional information. (yes/no) {\bf yes}
    \item The significance of any improvement or decrease in performance is judged using appropriate statistical tests (e.g., Wilcoxon signed-rank). (yes/partial/no) {\bf yes}
    \item This paper lists all final (hyper-)parameters used for each model/algorithm in the paper’s experiments. (yes/partial/no/NA) {\bf yes}
    \item This paper states the number and range of values tried per (hyper-) parameter during development of the paper, along with the criterion used for selecting the final parameter setting. (yes/partial/no/NA) {\bf yes}
\end{itemize}
\end{document}